\documentclass{article}

\usepackage{a4}
\usepackage{graphicx}
\usepackage{array}
\usepackage{hyperref}
\usepackage{layout}

\begin{document}

\begin{center}
  \Large{Towards optimization of quantum circuits}\\
  \renewcommand{\thefootnote}{\fnsymbol{footnote}}
  \normalsize \textit{Michal Sedl\'{a}k$^{1,2,}$\footnote[1]{michal.sedlak@savba.sk} and Martin Plesch$^{1,2}$\\
    $^{1}$Research Center for Quantum Information, Slovak Academy of Sciences,\\
    D\'{u}bravsk\'{a} cesta 9, 845 11 Bratislava, Slovak
    Republic,}

    \textit{$^{2}$QUNIVERSE, L\'{i}\v{s}\v{c}ie \'{u}dolie 116, 841 04 Bratislava, Slovakia}
\end{center}

Any unitary operation in quantum information processing can be
implemented via a sequence of simpler steps –- quantum gates.
However, actual implementation of a quantum gate is always
imperfect and takes a finite time. Therefore, seeking for a short
sequence of gates - efficient quantum circuit for a given
operation, is an important task. We contribute to this issue by
proposing optimization of the well-known universal procedure
proposed by Barenco et.al [1]. We also created a computer program
which realizes both Barenco's decomposition and the proposed
optimization. Furthermore, our optimization can be applied to any
quantum circuit containing generalized Toffoli gates, including
basic quantum gate circuits.

\section{Introduction}
Practical realization of quantum information processing requires
ability to prepare a quantum system in a chosen state, to perform
the desired operation on it and to read out the outcome via a
measurement. These tasks can be carried out on a collection of
two-level quantum systems - qubits. Since it is very difficult
(practically impossible) to properly control simultaneous
interaction among many qubits, the desired operation is usually
performed as a sequence of simpler operations - quantum gates. The
gate is accomplished by a temporal evolution of a system during
which only few qubits interact simultaneously. The complicated
operation is then built up by a sequence of quantum gates -
quantum circuit containing experimentally feasible gates. In the
quantum circuit model [2] the system of qubits is described as a
closed quantum system. Therefore the time evolution is unitary, and
each quantum gate is a unitary operator.

A set of (experimentally realizable) gates is called quantum gate
library. In the rest of the paper we work with the basic-gate
library [1], which contains all one qubit rotations and the
Controlled-NOT (CNOT) gate. This library is universal in the sense
that any unitary operation can be exactly achieved by a quantum
circuit containing only finite number of gates from the basic-gate
library. This universality was shown in 1995 constructively by
Barenco et. al. [1]. Since that time, much effort has been made to
propose a universal technique for finding an efficient quantum
circuit for a general unitary operation. Many research groups
focused on searching for an universal n-qubit circuit containing
the lowest possible number of CNOT gates (see e.g. [3],[4],[5]).
That means a circuit capable to achieve any unitary operator by
tuning the circuit's one-qubit gates. The number of CNOT gates is
important both, for its relation to the execution time of the
circuit, and also from the point of view of complexity of its
experimental implementation.

Shende, Markov and Bullock in [6] showed by dimension-counting arguments
that universal n-qubit circuits have to contain at least
$\frac{1}{4}(4^n-3n-1)$ CNOT gates. Although universal circuits
with the lowest number of CNOTs are known for the special case of
2 qubits (see Refs. [6],[7],[8]), for higher number of qubits it
remains an open problem. Furthermore, we have no guarantee that
the tuning of single qubit gates corresponding to an operator
realizable efficiently will lead to a straightforward
simplification of the universal circuit. Therefore, universal
n-qubit circuits often contain exponential number of CNOT gates
(with respect to n) also for n-qubit operations realizable by a
polynomial number of CNOTs. Intricacy of the simplification
(optimization) of universal circuits for a chosen operator can be
seen in the case of two qubit operators where the circuits with
the lowest number of CNOT gates for a given operator were found by
other means. On the other hand, for more than two qubits, the
universal circuits are the only standard approach to find a
quantum circuit for an arbitrary given operator.

In the present paper we show a simplification of the universal
n-qubit circuit proposed by Barenco et.al. [1]. We tried to find,
for an arbitrary given unitary operator, a quantum circuit
containing the lowest possible number of CNOT gates. Barenco's
decomposition utilizes the decomposition of unitary matrix into
multiplication of a diagonal matrix and two-level matrices. After
this decomposition one obtains a preliminary quantum circuit
containing generalized Toffoli gates. These gates will be finally
implemented by other constructions using basic quantum gates.

We examined some natural questions concerning generalized Toffoli
gates and we propose an optimization algorithm which combines the
found properties. This optimization algorithm can be applied to
any quantum circuit containing generalized Toffoli gates including
circuits containing basic gates. Hence, we can utilize our
optimization in several stages of Barenco's decomposition. To
perform Barenco's decomposition and the proposed optimization, we
created a computer program (the program can be downloaded from
www.quniverse.sk/people/sedlak/), which was used to estimate the
efficiency of the optimization.

The rest of the paper is organized as follows. We start with a
definition of the generalized Toffoli gate, which is followed by a
brief sketch of Barenco's et. al. decomposition. More details
about the procedure can be found in Refs. [1] and [10]. In section
3, we examine the properties of generalized Toffoli gates which
are combined to create an optimization algorithm in section 4. The
results obtained by the optimization algorithm are summarized in
the section 5.
\section{Preliminaries}
\subsection{Definition of Generalized Toffoli gate $\wedge_m (A)$}
The generalized Toffoli gate $\wedge_m (A)$ is an $(m+1)$--qubit
gate with $m$ control qubits $j_1,\ldots,j_m$, and one target
qubit $j_0$. The action of the gate on computational basis vectors
reads:
\begin{equation}
|x_1,\ldots,x_{j_0},\ldots,x_n \rangle \rightarrow
|x_1,\ldots\rangle \otimes A^{x_{j_1}\wedge \ldots \wedge x_{j_m}}
|x_{j_0} \rangle \otimes |\ldots,x_n \rangle , \label{rov1}
\end{equation}
where A is an operator (2x2 matrix) acting on one qubit. Thus the
target qubit of computational--basis vector is affected by the
operator A only if all control qubits are in the state
$|1\rangle$.

\subsection{Brief sketch of the Barenco's decomposition} The procedure can be
divided into four steps:
\begin{figure}
\begin{center}
\includegraphics[width=1.0\textwidth]{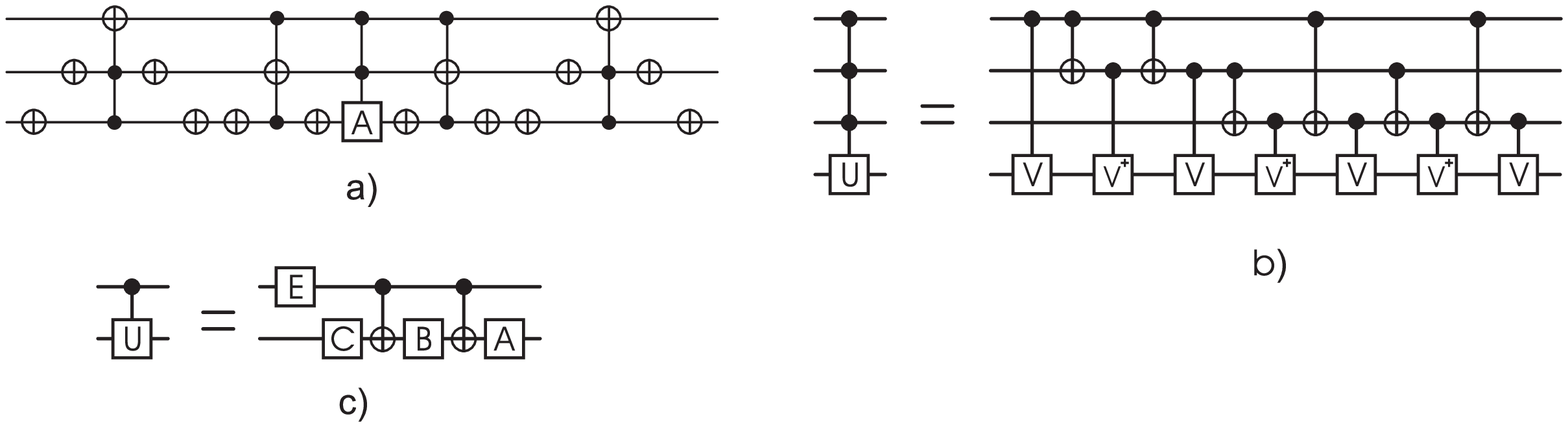}
\caption{Examples from parts of A. Barenco et. al. decomposition.}
\label{obr0}
\end{center}
\end{figure}

\begin{itemize}
\item{ \textbf{Step 1 -– QR decomposition.} Matrix of the chosen operator $U$
is written as:
\begin{equation}
U=D^{-1}.T^{-1}_{2,1}.T^{-1}_{3,1}.T^{-1}_{3,2}\ldots
T^{-1}_{2^n,2^n -2}.T^{-1}_{2^n,2^n -2},
 \label{rov0}
\end{equation}
where $D$ is a diagonal phase-matrix and $T_{pq}$ are matrices
acting nontrivially only in two--dimensional subspaces, which are
given by pairs of computational--basis vectors.}

\item{\textbf{Step 2 –- Decomposition of matrices $T_{pq}$ and $D$
into generalized Toffoli gates.} General matrix $T_{pq}$ operates on one pair of distinct
computational--basis vectors. However, generalized Toffoli gate $\wedge_{n-1} (A)$ changes one pair of computational--basis vectors which differ only in one qubit. Therefore, we first use $\wedge_{n-1} (\sigma_x)$ gates and NOT gates to perform a permutation which takes the pair of computational--basis vectors given by $T_{pq}$ to vectors differing in only one qubit. Then we apply the appropriate $\wedge_{n-1} (A)$ gate (together with some NOT gates) and finally undo the permutation. This allows to implement every matrix $T_{pq}$. To build the entries of diagonal matrix $D$, we use $\wedge_{n-1} (diag(.,.))$ gates surrounded by pairs of NOT gates. As an example, in Figure \ref{obr0}.a one can see the decomposition of matrix $T_{8,1}$ operating between vectors $|000\rangle$, $|111\rangle$ .}

\item{\textbf{Step 3 –- Simplification of generalized Toffoli
gates.} Gates $\wedge_{n-1} (A)$ are implemented by Controlled
1-qubit gates (less complicated generalized Toffoli gates
$\wedge_1 (V)$). As an example, Figure \ref{obr0}.b shows the
simplification of the $\wedge_3 (U)$ gate.}

\item{\textbf{Step 4 –- Decomposition of $\wedge_1 (V)$ gates
into basic quantum gates.} This step finishes the decomposition by
using only basic quantum gates in the circuit. The example in
Figure \ref{obr0}.c shows the worst case decomposition of the
$\wedge_1 (V)$ gate into four 1-qubit gates and two CNOT gates.
For some $\wedge_1 (V)$ fewer gates suffice.}
\end{itemize}

\section{Properties of generalized Toffoli gates}
\begin{figure}
\begin{center}
\includegraphics[width=1.0\textwidth]{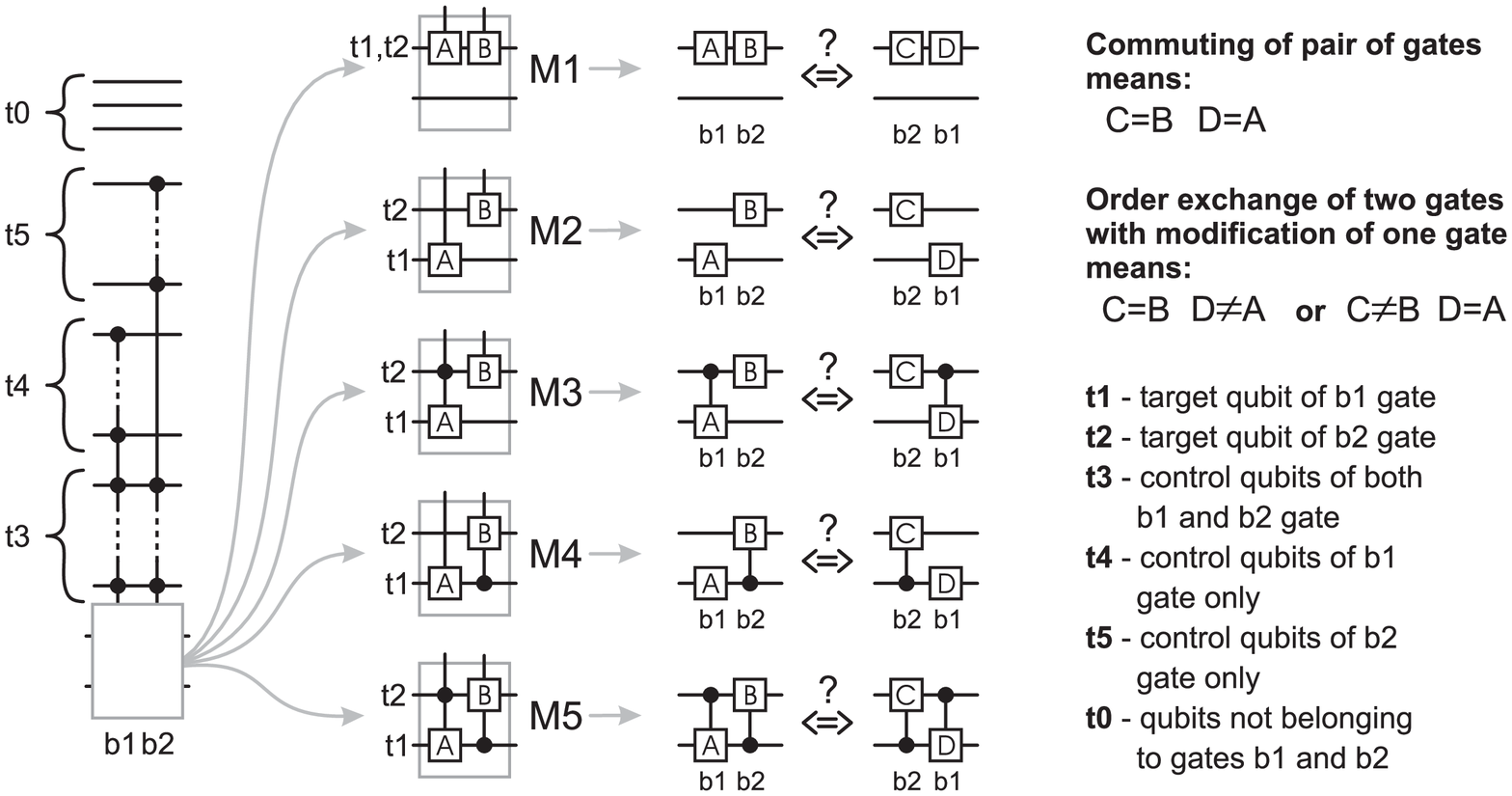}
\caption{Examining properties of generalized Toffoli gates}
\label{obr1}
\end{center}
\end{figure}

By renaming qubits, each pair of generalized Toffoli gates can be
drawn and denoted as shown in Figure \ref{obr1}. Since every
linear operator is fully defined through it's action on basis
vectors, we will show equality of two operators via their action
on computational--basis vectors. In what follows, the
action of the gates b1,b2 is always described with respect to the
computational basis.
\subsection{Commutativity of pair of gates}
Qubits of the type t0, t3, t4 and t5 (see definition in Figure
\ref{obr1}) do not affect commutativity of the gates $b1=\wedge_k
(A)$ and $b2=\wedge_l (B)$, because they control the action on
qubits of the type t1, t2 in the same way in both orderings. From
this property it follows that it suffices to examine the cases M1
- M5. In each of these cases the action of the gates on
computational--basis vectors in both orderings yields four
equations, which give constraints on entries of the 2x2 matrices
$A$ and $B$. In the case M1 it is obvious that the gates b1 and b2
commute if matrices A and B commute. In the case M2 the gates
surely always commute. In the case M3 gates b1 and b2 commute if
the matrix B is diagonal. In the case M4 the gates commute if the
matrix A is diagonal. Thus we see that in the cases M3, M4 the gates
b1 and b2 commute if diagonal matrix B (resp. A) is passing
through the control qubit of neighbouring gate. The situation in
the case M5 is a bit different, and finally it turns out that there
are three ways how to fulfill the aforementioned equations: i)
Both matrices A and B are diagonal, ii) $A=diag(e^{\imath
\alpha},1)$ and no constraint on $B$, iii) $B=diag(e^{\imath
\beta},1)$ and no constraint on $A$.
\subsection{Exchange of two gates with modification of one gate}
We consider only modification of one--qubit operator(2x2 matrix)
from the definition of the generalized Toffoli gate (\ref{rov1}).
We can consider the gates as not commuting, because otherwise it follows
from the unitarity that neither of the gates can be modified.
Let's look at exchange of gates $b1=\wedge_k (A)$, $b2=\wedge_l
(B)$ for gates $b2n=\wedge_l (C)$, $b1=\wedge_k (A)$. The
previous argument tells us that $B\neq C$. If the gate b1 had
qubits t4, then there would exist a computational--basis vector
with at least one qubit t4 in the state $|0\rangle$ and all qubits
t1, t3, t5 in state $|1\rangle$ which would reveal the difference
between B and C, i.e. difference between the gates b2 and b2n. Thus
if we want this exchange to be possible, the gate b1 must not
involve qubits t4. Gates can have the other types of qubits,
because they do not enable separate action of the gate b2, and
b2n. Similarly, as in the case of the commutativity of gates, it
remains to examine the cases M1, M3 -- M5 and to solve similar
equations. The results of these technical calculations are presented in
Table \ref{tabulka1}.
\begin{table}[htb]
\begin{center}
\begin{tabular}{|l|l|l|}
\hline
Case &Constraints on A, B& Corresponding C \\
\hline
\includegraphics[width=0.25\textwidth]{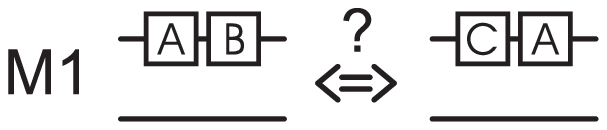}&no constraints & $C=A^\dag$.B.A\\
\hline
\includegraphics[width=0.25\textwidth]{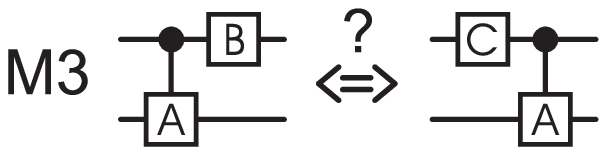}&$A=
\left(
 \begin{array}{cc}
 e^{\imath\varphi}&0 \\
 0&e^{\imath\varphi}\\
  \end{array}
\right),
 B=
\left(
 \begin{array}{cc}
 B_{00}&B_{01} \\
 B_{10}&B_{11}\\
  \end{array}
\right) $&
$C=
\left(
 \begin{array}{cc}
 B_{00}&B_{01} e^{\imath\varphi} \\
 B_{10}  e^{-\imath\varphi}&B_{11}\\
  \end{array}
\right)$\\
\hline
\includegraphics[width=0.25\textwidth]{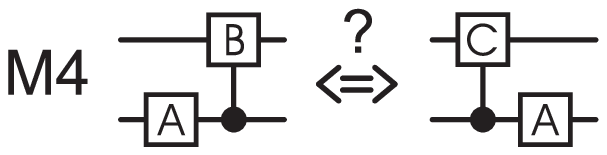}&It is not possible to fulfill the equations &\\
\hline
\includegraphics[width=0.25\textwidth]{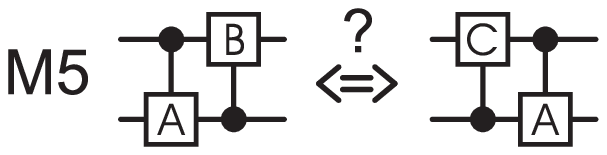}&$A=
\left(
 \begin{array}{cc}
 e^{\imath\varphi_1}&0 \\
 0&e^{\imath\varphi_2}\\
  \end{array}
\right),
 B=
\left(
 \begin{array}{cc}
 B_{00}&B_{01} \\
 B_{10}&B_{11}\\
  \end{array}
\right) $&
$C=
\left(
 \begin{array}{cc}
 B_{00}&B_{01} e^{\imath\varphi_2} \\
 B_{10}  e^{-\imath\varphi_2}&B_{11}\\
  \end{array}
\right)$\\
\hline
\end{tabular}
\end{center}
\caption{Exchange of gates $b1=\wedge_k (A)$, $b2=\wedge_l (B)$
for gates $b1=\wedge_l (C)$, $b2=\wedge_k (A)$, which is
possible only if there are no qubits t4 (see Figure \ref{obr1}).}
\label{tabulka1}
\end{table}
For the completeness, the conditions for exchange of gates
$b1=\wedge_k (A)$, $b2=\wedge_l (B)$ for gates $b2=\wedge_l
(B)$, $b1n=\wedge_k (D)$ (completely analogous to the previous
one) are summarized in Table \ref{tabulka2}.

\begin{table}[htb]
\begin{center}
\begin{tabular}{|l|l|l|}
\hline
Case &Constraints on A, B& Corresponding D \\
\hline
\includegraphics[width=0.25\textwidth]{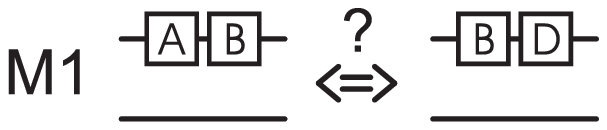}&no constraints & $D=B^\dag.A.B$\\
\hline
\includegraphics[width=0.25\textwidth]{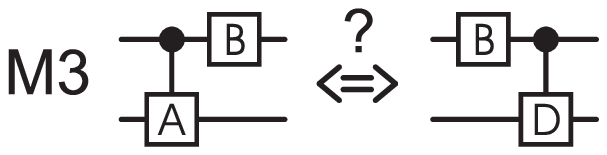}&It is not possible to fulfill the equations &\\
\hline
\includegraphics[width=0.25\textwidth]{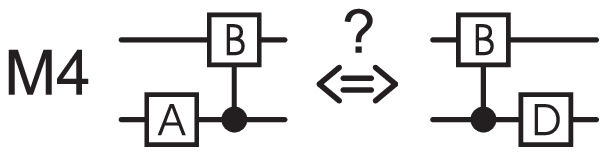}&$B=
\left(
 \begin{array}{cc}
 e^{\imath\varphi}&0 \\
 0&e^{\imath\varphi}\\
  \end{array}
\right),
 A=
\left(
 \begin{array}{cc}
 A_{00}&A_{01} \\
 A_{10}&A_{11}\\
  \end{array}
\right) $&
$D=
\left(
 \begin{array}{cc}
 A_{00}&A_{01} e^{-\imath\varphi} \\
 A_{10} e^{\imath\varphi}&A_{11}\\
  \end{array}
\right)$\\
\hline
\includegraphics[width=0.25\textwidth]{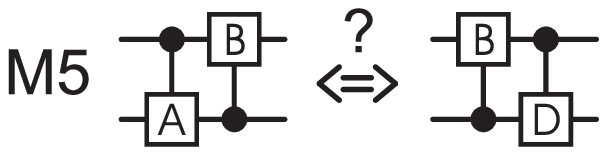}&$B=
\left(
 \begin{array}{cc}
 e^{\imath\varphi_1}&0 \\
 0&e^{\imath\varphi_2}\\
  \end{array}
\right),
 A=
\left(
 \begin{array}{cc}
 A_{00}&A_{01} \\
 A_{10}&A_{11}\\
  \end{array}
\right) $&
$D=
\left(
 \begin{array}{cc}
 A_{00}&A_{01} e^{-\imath\varphi_2} \\
 A_{10} e^{\imath\varphi_2}&A_{11}\\
  \end{array}
\right)$\\
\hline
\end{tabular}
\end{center}
\caption{Exchange of gates $b1=\wedge_k (A)$, $b2=\wedge_l (B)$
for gates $b1=\wedge_l (B)$, $b2=\wedge_k (D)$, which is
possible only if there are no qubits t5(see Figure \ref{obr1}).}
\label{tabulka2}
\end{table}

 \subsection{Conditions for merging two gates into one}
\begin{figure}
\begin{center}
\includegraphics[width=1.0\textwidth]{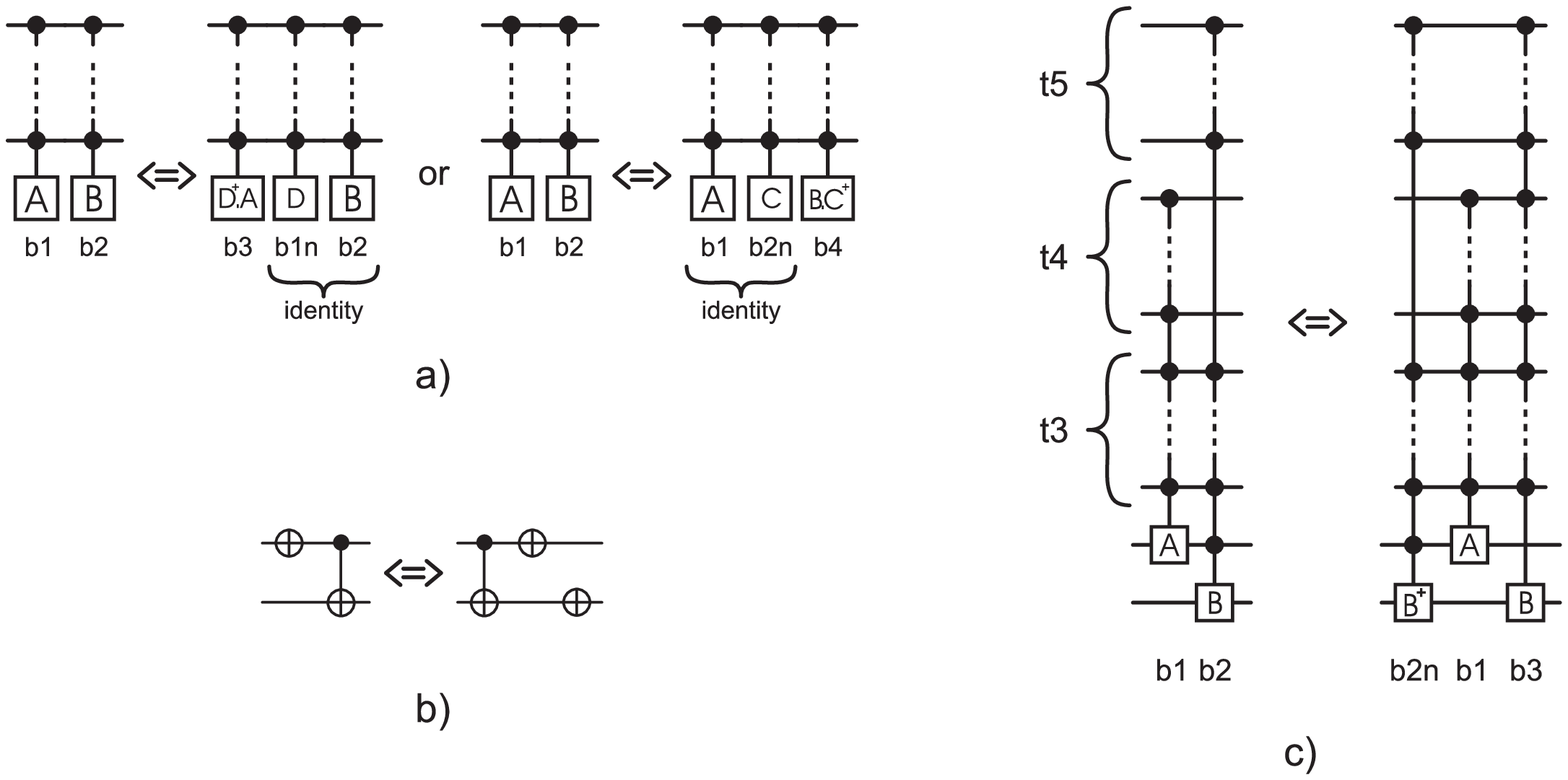}
\caption{Circuit identities. The one from c) holds if matrix A has
vanishing diagonal elements.} \label{obr2}
\end{center}
\end{figure}

First of all, we will examine the circumstances under which two
generalized Toffoli gates form an identity. This will enable us to
formulate conditions of merging two generalized Toffoli gates into
one. We consider the gates $b1=\wedge_k (A)$ and $b2=\wedge_l
(B)$, both different from the identity. The gates b1 and b2 must
not involve the qubits t4 and t5, because they involve action of
either the gate b1 or b2 on some subspace of the Hilbert space,
where we see their difference from the identity. It suffices to
examine the cases M1 -- M5 (see Figure \ref{obr1}), because the
gates b1 and b2 act nontrivially only on computational-basis
vectors with all qubits of the type t3 in the state $|1\rangle$
and do not modify the qubits other than t1 and t2. So in each case
we write down the transformation carried out by the gates b1, b2
and require it to be the identity. The resulting constraints on
elements of the matrices A and B are presented in Table
\ref{tabulka3}.
\begin{table}[htb]
\begin{center}
\begin{tabular}{|l|l|}
\hline
Case &Constraints on A, B\\
\hline
M1&$A.B=1$ \\
\hline
M2&$A=1.e^{\imath\varphi} , B=1.e^{-\imath\varphi} $ \\
\hline
M3&$A=diag(e^{-\imath\varphi},e^{-\imath\varphi}) , B=diag(1,e^{\imath\varphi})$ \\
\hline
M4&$A=diag(1,e^{\imath\varphi}), B=diag(e^{-\imath\varphi},e^{-\imath\varphi}) $ \\
\hline
M5&$A=diag(1,e^{\imath\varphi}), B=diag(1,e^{-\imath\varphi})$ \\
\hline
\end{tabular}
\end{center}
\caption{Two generalized Toffoli gates $b1=\wedge_k (A)$ and
$b2=\wedge_l (B)$ form the identity if they don't have qubits of
the type t4, t5 (see Figure \ref{obr1}) and fulfill these
constraints.} \label{tabulka3}
\end{table}
Obviously, if we have two neighbouring generalized Toffoli gates
which form identity, we remove them from the circuit. Also, if we
have two such neighbouring gates $b1=\wedge_k (A)$ and
$b2=\wedge_l (B)$, which do not form the identity only because
either of the matrices A or B does not fulfill the constraints
from Table \ref{tabulka3}, it is possible to simplify the circuit.
It suffices to suitably divide the gate b1 or b2 as shown in
Figure \ref{obr2}a) and to remove the pair of gates forming the
identity. This finally leads us to merging the two gates into one.
\subsection{Exchange of two gates with help of one additional gate}
During the Barenco's decomposition we work with circuits, where
NOT gates act on control qubit of a neigbouring generalized
Toffoli gate $\wedge_m (\sigma_X)$. These pairs of gates do not
commute and their order cannot be exchanged even when we modify
one of them. This is the reason why we generalized the well
known CNOT identity shown in Figure \ref{obr2}.b. Our
generalization is depicted in Figure \ref{obr2}.c, where matrices
A, B are not arbitrary, but restricted as we will describe below.
In the case there are no qubits t4, the requirement of equality of
transformations performed by circuits from Figure \ref{obr2}.c
only tells us that the matrix A must have vanishing diagonal
elements. But if qubits t4 are present, then we must fulfill also
the condition $B=B^\dag$, because only gates b2, and b2n act on
computational--basis vectors with at least one qubit t4 in the state
$|0\rangle$.
\section{Optimization algorithm}
The main idea of our approach to an optimization of quantum
circuits containing generalized Toffoli gates is based on using
the properties of $\wedge_k (.)$ gates described above. We drag
the selected gate to the left until we merge it with a
neighbouring gate or we are not able to drag it further. After
that, we try the same to the right and then select another gate, and
repeat the whole procedure. We do these steps, until the number of
gates in the circuit is decreasing. We perform the dragging as
follows: If gates commute we exchange them, otherwise we try to
exchange them while modifying one of them. If it is not possible
to use any of these options, we use exchange of two gates with the
help of one additional gate, but only if there are no qubits t4
(see Figure \ref{obr2}.c). In addition, we use it only once before
we succeed in merging the gates. This guarantees that the number
of gates in the circuit will not grow. In case we use an
exchange of two gates with the help of one additional gate, the
number of gates in the circuit will be the same, but one
generalized Toffoli gate will have less control qubits, which
leads to less basic gates in subsequent steps of Barenco's
decomposition.
\section{Results}
Due to the fact that Barenco's et. al. decomposition [1] leads to
quantum circuits containing hundreds of gates even in the case of
3-qubit operations, it is not possible to perform the
decomposition and the proposed optimization by hand. This reason
stimulated us to create a computer program which realizes both
Barenco's decomposition and optimization algorithm proposed in
this paper. We can show analytically that, in the worst case of a
2-qubit operation, our optimization improves Barenco's
decomposition to obtain circuit containing 10 CNOTs instead of 20
CNOTs. However, it is known that three CNOTs suffice to implement
any 2-qubit operation, therefore we see that the proposed
optimization is only a partial one. We examined the
efficiency of our optimization for different subsets of operators
acting on various numbers of qubits numerically. Our approach
starts by randomly generating the operator with known upper bound
on number of CNOTs needed for its implementation. This is done by
computing the operation corresponding to a circuit build from that
number of CNOTs and randomly picked 1-qubit gates. Then we
decompose this operator by Barenco's et. al. decomposition
with/without our optimization. We did this several times and
evaluated the results. One would expect that the number of CNOTs
in the circuit created by Barenco's decomposition with our
optimization will strongly depend on the operator we are
decomposing. We have found out that, except for the operators
generated by circuits containing artificially chosen 1-qubit
gates, each decomposition with the proposed optimization leads to
a circuit with exactly the same number of CNOT gates (for the
chosen number of qubits). This is caused by a redundancy in the
blocks of generalized Toffoli gates, which our optimization is not
able to remove in generic cases. For a more quantitative overview
see Table \ref{tabulka4}. This table also shows comparison with
the NQ [4] and the CS [5] decompositions, which are the best
performing universal decompositions in the worst case of unitary
operators. The asymptotic number of CNOT gates used by Barenco's
decomposition to implement any n--qubit unitary is O($n^3 4^n$).
Our numerical investigation suggests that the asymptotics may be
the same also with using the proposed optimization. The NQ and CS
decompositions are more efficient in general (creating roughly
$1/2 \times 4^n$ CNOT gates in the worst case of a unitary
operator), because their procedure contains steps which
systematically remove a part of the redundancy introduced in
previous steps.
\begin{table}
\begin{center}
\begin{tabular}{|c|c|c|c|c|}
\hline
& \multicolumn{4}{|c|}{Decomposition}\\
\hline
Number of qubits & A. Barenco's & A.Barenco's & NQ & CS \\
n &  & \textbf{+ Our optimization} &  &  \\ \hline\hline 2 &
\multicolumn{1}{|r|}{20} & \multicolumn{1}{|r|}{10} &
\multicolumn{1}{|r|}{3} & \multicolumn{1}{|r|}{4} \\ \hline 3 &
\multicolumn{1}{|r|}{576} & \multicolumn{1}{|r|}{379} &
\multicolumn{1}{|r|}{21} & \multicolumn{1}{|r|}{26} \\ \hline 4 &
\multicolumn{1}{|r|}{8000} & \multicolumn{1}{|r|}{6278} &
\multicolumn{1}{|r|}{105} & \multicolumn{1}{|r|}{118} \\ \hline 5
& \multicolumn{1}{|r|}{91520} & \multicolumn{1}{|r|}{76208} &
\multicolumn{1}{|r|}{465} & \multicolumn{1}{|r|}{494} \\ \hline
\end{tabular}%
\end{center}
\caption{The Number of CNOT gates in the circuits produced by
various decompositions for the generic unitary operator.}
\label{tabulka4}
\end{table}

\section{Summary}
Finding an efficient quantum circuit for a given n-qubit operation
is an important task in the quantum circuit model of computation.
Few universal procedures performing this task were proposed, but
their efficiency (the number of  created CNOT gates) is very often
known only for the worst case of n-qubit unitary operators.
However, it is believed that interesting operations might require
only polynomial number of CNOT gates with respect to the number of
qubits. Hence, it is very important to know the performance of
such universal procedures on this kind of operators. Therefore, we
proposed an optimization of the universal procedure by
Barenco et. al., and examined it's efficiency on the operators
realizable by a small number of CNOT gates. To perform this task
we created a computer program performing Barenco's procedure
together with the proposed optimization. The results show that
this procedure is in general not as efficient as the NQ and CS
decompositions and still leaves some redundancy in the created
circuit. On the other hand, our optimization is not restricted to
be used only with Barenco's decompositon and can be aplied on any
quantum circuit containing generalized Toffoli gates which include
circuits containing basic quantum gates. This can be useful once
we have some basic gate circuit corresponding to unitary operator
we are decomposing. Our optimization can also be useful in
situations when the quantum algorithm is given as a sequence of
efficient sub-circuits performing the sub-tasks. A very similar
idea to our optimization was proposed and extended to a slightly
more general framework by D. Maslov, G. W. Dueck and D.M. Miller
in [11]. It's not possible to correctly compare their numerical
results to ours, since they work with different gate
library containing one--qubit gates, controlled--NOT gate, and
controlled-sqrt-of-NOT gate. But roughly we can say that the
portion of the gates removed in the particular examples they
present is very similar to the portion of gates our optimization
removes in the case of Barenco's decomposition. All published
optimizations of quantum circuits are based on exchanging
sequences of gates for shorter ones doing precisely the same
thing. To propose a better optimization strategies it seems that
we probably need to understand more deeply what is computed in the
considered part of the circuit.

\section*{ACKNOWLEDGMENTS}
This work was supported by the European Union projects QAP,
CONQUEST and by the projects APVT-99-012304, INTAS 04-77-7289.
\\[1cm]

%%%A hand-made bibliography example you can use
\noindent[1] A. Barenco et.al.,"Elementary gates for quantum
computation", Physical Review A, March 22, 1995 (AC5710)

\noindent[2] D. Deutsch, "Quantum computational networks", Proc.
R. Soc. London A 425, 73 (1989).

\noindent[3] M. M\"{o}tt\"{o}nen, J. Vartiainen,V. Bergholm and M.
Salomaa,"Quantum Circuits for General Multiqubit Gates", Physical
Review Letters \textbf{93}, 130502 (2004)

\noindent[4] V. Shende, S. Bullock and I. Markov, "Synthesis of Quantum Logic Circuits",
IEEE Transactions on Computer-Aided Design \textbf{25} no. 6 pg. 1000 (2006)

\noindent[5] V. Bergholm, J. Vartiainen, M. M\"{o}tt\"{o}nen and
M.Salomaa, "Quantum circuits with uniformly controlled one-qubit
gates", Phys. Rev. A \textbf{71}, 052330(2005)

\noindent[6] V. Shende, I. Markov and S. Bullock, "Minimal
Universal Two-Qubit controlled-NOT-based Circuits", Physical
Review A \textbf{69}, 062321 (2004)

\noindent[7] F. Vatan and C. Williams, "Optimal Quantum Circuits
for General Two-Qubit Gates", Physical Review A \textbf{69},
032315 (2004)

\noindent[8] G. Vidal and C.Dawson, "Universal quantum circuit for
two-qubit transformations with three controlled-NOT gates", Phys.
Rev. A \textbf{69}, 010301(2004)

\noindent[9] V. Shende, S. Bullock and I. Markov, "Recognizing
small-circuit structure in two-qubit operators", Physical Review A
\textbf{70}, 012310 (2004)

\noindent[10] M. Nielsen and I. Chuang, "Quantum Computation and
Quantum Information", Cambridge University Press (2000)

\noindent[11] D. Maslov, G. W. Dueck and D.M. Miller, "Quantum
Circuit Simplification and Level Compaction", Proceedings of the
conference on Design, Automation and Test in Europe - Volume 2
1208 - 1213 (2005), arXiv:quant-ph/0604001 (2004)

\end{document}